# Reaction of the NAD(P)H : Flavin Oxidoreductase from *Escherichia coli* with NADPH and Riboflavin : Identification of Intermediates [†]

Vincent Nivière [‡]*, Maria A. Vanoni [§],

Giuliana Zanetti [§] and Marc Fontecave [‡]

[‡] Laboratoire de Chimie et Biochimie des Centres Redox Biologiques,

DBMS-CEA / CNRS / Université Joseph Fourier, 17 Avenue des Martyrs,

38054 Grenoble, Cedex 9, France

[§] Dipartimento di Fisiologia e Biochimica Generali, Universita' degli Studi di

Milano, Via Celoria 26, I-20133 Milano, Italy

Running Title : Reaction intermediates of Fre

* To whom correspondence should be addressed. Telephone : 33-(0)4-76-88-91-09.  Fax : 33-(0)4-76-88-91-24. E-mail : niviere@cbcrb.ceng.cea.fr

[1] Abbreviations : Fre, NAD(P)H : flavin oxidoreductase; FNR, ferredoxin-NADP$^+$ reductase; PDR, phthalate dioxygenase reductase; PDR-FeS, truncated form of phthalate dioxygenase reductase that lacks the iron-sulfur domain; Rf, riboflavin; MC, Michaelis complex; CT, charge-transfer complex.

ABSTRACT

Flavin reductase catalyzes the reduction of free flavins by NAD(P)H. As isolated, *Escherichia coli* flavin reductase does not contain any flavin prosthetic group, but accommodates both the reduced pyridine nucleotide and the flavin substrate in a ternary complex prior to oxidoreduction. The reduction of riboflavin by NADPH catalyzed by flavin reductase has been studied by static and rapid kinetics absorption spectroscopies. Static absorption spectroscopy experiments revealed that, in the presence of riboflavin and reduced pyridine nucleotide, flavin reductase stabilizes, although to a small extent, a charge-transfer complex between $NADP^+$ and reduced riboflavin. In addition, reduction of riboflavin was found to be essentially irreversible. Rapid kinetics absorption spectroscopy studies demonstrated the occurrence of two intermediates with long-wavelength absorption during the catalytic cycle. Such intermediate species exhibit spectroscopic properties similar to those of charge-transfer complexes between oxidized flavin and NAD(P)H, and reduced flavin and $NAD(P)^+$, respectively, which have been identified as intermediates during the reaction of flavoenzymes of the ferredoxin-$NADP^+$ reductase family. Thus, a minimal kinetic scheme for the reaction of flavin reductase with NADPH and riboflavin can be proposed. After formation of the Michaelis complex of flavin reductase with NADPH and riboflavin, a first intermediate, identified as a charge-transfer complex between NADPH and riboflavin, is formed. It is followed by a second charge-transfer intermediate between enzyme-bound NADP$^+$ and reduced riboflavin. The latter decays yielding the Michaelis complex of flavin reductase with NADP$^+$ and reduced riboflavin, which then dissociates to complete the reaction. These results support the initial hypothesis of a structural similarity of flavin reductase to the enzymes of the ferredoxin-NADP$^+$ reductase family and extend it at a functional level.

Flavins are well known as prosthetic groups of flavoproteins. However, they can also serve as substrates of a class of enzymes named flavin reductases. These enzymes are defined by their ability to catalyze the reduction of free flavins (riboflavin, FMN or FAD) by reduced pyridine nucleotides, NADPH or NADH. Flavin reductases are present in all living organisms, from bacteria to mammals (*1*). Their physiological role is still unclear, even though there is indirect evidence for their function, at least in prokaryotes, in bioluminescence (*2*), ferric reduction (*3*), DNA synthesis (*4, 5*) or oxygen reduction (*6*). Recently, flavin reductases have also been found to be associated with oxygenases involved in desulfurization process of fossil fuels (*7*) and antibiotic biosynthesis (*8-10*).

Two classes of flavin reductases, which do not share any sequence similarities, have been found. Class I enzymes do not contain any flavin prosthetic group and cannot be defined as flavoproteins (*4*), whereas class II enzymes are canonical flavoproteins (*11, 12*).

The prototype of class I flavin reductases is an enzyme, named Fre [1], which was discovered in *Escherichia coli* as a component of a multienzymatic system involved in the activation of ribonucleotide reductase, a key enzyme for DNA biosynthesis (*4, 5*). Fre consists of a single polypeptide chain of 233 amino acids, with a molecular mass of 26,212 Da (*13*). The electron donor can be either NADPH or NADH and riboflavin is the best substrate (*4*).

Steady-state kinetic studies have shown that Fre functions through an ordered mechanism, with NADPH binding first. This indicates that the protein provides a site which accommodates both the reduced pyridine nucleotide and the flavin substrate in a ternary complex prior to electron transfer (*14*). Substrate specificity studies have shown that the isoalloxazine ring of the flavin is the only part of the molecule which is recognized by Fre (*14*).

On the basis of sequence similarities, it was proposed that Fre belongs to a large family of flavoenzymes of which spinach ferredoxin-NADP$^+$ reductase (FNR) is the structural prototype (*15, 16*). Several members of this family have been structurally characterized, including FNR from other species (*17, 18*), cytochrome P-450 reductase (*19*), cytochrome b5 reductase (*20*), nitrate reductase (*21*) and phthalate dioxygenase reductase (PDR) (*22*).

Although the overall sequence similarity is low, Fre contains a sequence of four amino acids, $R_{46}PFS_{49}$, similar to the $R_{93}LYS_{96}$ region of spinach FNR. The three-dimensional crystallographic structures of proteins of the FNR family (*15-22*) clearly show that these residues, which are highly conserved in all family members, are involved in the interaction between the protein and the flavin isoalloxazine ring. In FNR, Ser96 has been proposed to make a hydrogen bond to the N(5) position of the FAD cofactor (*15, 16*). This residue was found to be essential for activity by site-directed mutagenesis (*23*). The same approach allowed us to demonstrate that Ser49 of Fre (corresponding to Ser96 of FNR) is indeed involved in flavin binding, and that it is also essential for activity (*24*). In addition, as in FNR, we have shown that Ser49 of Fre interacts with the N(5) of the isoalloxazine ring of the flavin substrate (*24*).

On the basis of these observations, the hypothesis can be made that Fre and FNR-family members might share mechanistic features in spite of their different affinities for the flavin, which make Fre a flavin reductase and FNR-family members flavoproteins.

The mechanism of the reaction of several flavoenzymes of the FNR family with NAD(P)H has been previously studied under pre-steady state conditions in a stopped-flow apparatus (*23, 25-28*). In these studies, it was shown that the reduction of the flavin cofactor by NAD(P)H occurs in discrete steps and involves at least four reaction intermediates (Scheme 1) : (i) binding of NAD(P)H to the enzyme active site to form the Michaelis complex (MC-1), (ii) conversion of MC-1 to the charge-transfer complex between oxidized flavin and NAD(P)H (CT-1), (iii) reduction of the flavin through the transfer of a hydride anion from NAD(P)H C(4) position to the flavin N(5) atom and formation a second charge-transfer complex (CT-2) between reduced flavin and NAD(P)$^+$, (iv) decay of the latter to yield the Michaelis complex between reduced enzyme and NAD(P)$^+$ product (MC-2), which then dissociates to give the reduced enzyme and free NAD(P)$^+$. The four species are spectrally distinguishable and the absorption properties of CT-1 and CT-2 are well documented for enzymes of this class (*26*).

The rapid reaction methodology has been thus applied to the reaction catalyzed by Fre. We here report the detection of reaction intermediates, which we identified as enzyme-bound charge-transfer complexes between oxidized riboflavin and NADPH and between reduced riboflavin and $NADP^+$. These results further support the structural and functional relationship between Fre and the members of the FNR family.

MATERIALS AND METHODS

*Materials.* Riboflavin (Rf), NADPH, NADP$^+$, glucose, glucose 6-phosphate were purchased from Sigma. 5-deaza-5-carba-riboflavin (5-deazaflavin) was a gift of Dr. Dale E. Edmondson, Department of Biochemistry, Emory University, Atlanta, GA, USA. *Aspergillus niger* glucose oxidase and *Leuconostoc mesenteroides* glucose 6-phosphate dehydrogenase were from Sigma.

*Production of recombinant flavin reductase.* Purification of the recombinant *E.coli* Fre was done using a two-step protocol, Phenyl-Sepharose and Superdex 75 as previously reported (*14*). Purified Fre solutions were concentrated at 4°C using an Amicon Centricon-30 centrifugal concentrator.

*Enzyme Assay.* Flavin reductase activity was determined at 25 or 5°C from the decrease of the absorbance at 340 nm ($\varepsilon_{340}$ = 6.22 mM$^{-1}$ cm$^{-1}$) due to the oxidation of NADPH. NADPH concentration was determined spectrophotometrically. Under standard conditions, the cuvette contained, in a final volume of 1 ml, 50 mM Tris/HCl, pH 7.6 or 50 mM Hepes/KOH, pH 7.6, glycerol 10% (buffer A), 0.2 mM NADPH and 20 μM Rf. Steady-state kinetic measurements were performed by varying the concentration of NADPH at a fixed Rf concentration (20 μM). The reaction was initiated by adding 0.2 μg of enzyme. One unit of activity was defined as the amount of enzyme catalyzing the oxidation of 1 nmol of NADPH per min. Protein concentration was determined using the Bio-Rad Protein Assay reagent (*29*) and bovine serum albumin as a standard protein. When applicable, values are shown ± standard deviation.

*Static absorption spectroscopy.* Spectra were recorded in quartz cuvettes designed for anaerobic work (*30*), using a Hewlett-Packard diode array 8452A spectrophotometer interfaced to a Hewlett-Packard 89500A ChemStation. When necessary, enzyme and reagent solutions were made anaerobic by repeated cycles of evacuation and saturation with nitrogen that had been purified from residual oxygen on passage through a heated BASF catalyst (*30*). Conditions used during individual experiments are specified in the corresponding figure legends. Photochemical reduction (*31*) were performed in a bath maintained at 10°C.

*Rapid reaction kinetics.* Experiments were performed using a Hi-Tech SF-61 stopped-flow spectrophotometer interfaced with a Macintosh IIci computer. Anaerobiosis in the instrument syringes and plumbing was achieved by flushing them with an anaerobic solution of 10 mM dithionite in 50 mM Tris/HCl, pH 8.5, which was allowed to scavenge oxygen overnight and was substituted by anaerobic 25 mM Hepes/KOH, pH 7.5 containing 1 mM EDTA, 10% glycerol, 2 mM glucose and 10 U/ml glucose oxidase, 2 h prior to the beginning of the experiment. Enzyme solutions were made anaerobic in a tonometer by repeated cycles of evacuation and saturation with oxygen-free nitrogen (*30*). Substrate solutions were instead made anaerobic by bubbling oxygen-free nitrogen through them for at least 15 min. For each reaction equal volumes (70 µl) of enzyme and substrate solutions were rapidly mixed. The KISS program (Kinetic Instruments Inc., USA) was used for data acquisition and analysis. When needed, absorbance spectra of the reaction mixtures of the substrate solutions were recorded using the rapid scanning device of the instrument.

RESULTS AND DISCUSSION

*Spectrophotometric titration of riboflavin with Fre.*

In order to gather preliminary information on the absorbance properties of various species that may form during the reaction of Fre with NADPH and riboflavin, in a first set of experiments we studied the modification of the absorbance spectrum of Rf induced by addition of Fre. Previous work (32) has shown that binding of Rf to Fre can occur, as monitored by flavin fluorescence quenching. Aliquots of a Fre solution (521 µM) were added to a solution of Rf (11.7 µM) in 50 mM Hepes/KOH buffer, pH 7.6, 10% glycerol (buffer A), at 10°C. As shown in Figure 1, binding of Rf to Fre caused well defined absorbance changes. The two maxima in the visible region of the free Rf spectrum shifted to longer wavelengths, from 374 to 382 nm, and from 446 to 456 nm, respectively, with a decrease of the extinction coefficient which is more marked for the 446 nm peak. A plot of fractional absorbance changes at 446 nm as a function of the [Fre]/[Rf] ratio is shown in the inset of Figure 1. The data were fitted to Eq. 1 :

(1)  $(A_0-A_x)/(A_0-A_f) =$

$-\{-(K_d/[Fre]+[Fre]/[Rf]+n)+SQR[(K_d/[Fre]+[Fre]/[Rf]+n)^2-4n]\}/2$

where : $A_0$ , initial absorbance; $A_x$ , absorbance after a given addition of

Fre; $A_f$, absorbance at the end of the titration; n, number of Rf binding sites per Fre molecule; $K_d$, dissociation constant of the Fre-Rf complex. Using this equation, a 1:1 stochiometry for Rf binding to Fre was confirmed (n = 1.16 ±0.08) and a $K_d$ of (3.6 ±1) µM was calculated. The $K_d$ value was comparable to that determined previously by flavin fluorescence quenching experiments (6 µM, *32*).

At the highest Fre concentration used in the experiment, it was calculated from Eq. 1 that 86.4% Rf was present in complex with Fre. From these data, it was possible to determine the extinction coefficient of the Fre-Rf complex ($\epsilon_{Fre-Rf}$) at a given wavelength, using Eq. 2 :

$$A = \epsilon_{Fre-Rf} * [Fre-Rf] + \epsilon_{Rf} * [Rf] \qquad (2)$$

where : A, absorbance at a given wavelength; $\epsilon_{Fre-Rf}$, extinction coefficient of the Fre-Rf complex at the given wavelength; $\epsilon_{Rf}$, extinction coefficient of free Rf at this wavelength. The extinction coefficients of Fre-Rf were calculated to be 10.7 and 11.5 $mM^{-1}$ $cm^{-1}$ at 446 and 456 nm respectively.

In order to study the influence of the pyridine nucleotide on the dissociation constant and the light absorption spectrum of the Fre-Rf complex, $NADP^+$ and 3-amino pyridine adenosine dinucleotide phosphate (AADP) were used as NADPH structural analogs in equilibrium experiments.

NADPH itself could not be used as it is the second substrate of the enzyme and reacts with Rf efficiently. No absorbance changes were observed when up to 21-fold excess NADP$^+$ or 30-fold AADP were added to solutions containing Fre (29 µM) and Rf (10.3 µM). Furthermore, the presence of AADP (166 µM) had no effect on the type of absorbance changes observed during the titration of Rf (11.7 µM) with a solution of Fre (521 µM) nor on the calculated dissociation constant of the Fre-Rf complex (Inset of Fig. 1). The lack of effect of NADP$^+$ and AADP on the interaction between Fre and Rf is in keeping with the fact that NADP$^+$ is a poor inhibitor of Fre ($K_i$ = 5 mM, *14*) and that AADP exerts no inhibition on Fre reaction at concentrations as high as 2 mM (this work).

*Fre-catalyzed riboflavin reduction by NADPH.*

In order to study the absorbance changes associated with the reaction of Fre with NADPH and Rf, aliquots of a 4.21 mM solution of NADPH were added to a solution containing Rf (50 µM) and Fre (175 µM) in buffer A, under anaerobiosis. Absorbance changes were monitored spectrophotometrically. As calculated from the $K_d$ value of 3.6 µM determined for the Fre-Rf complex, under these conditions, approximately 97% Rf was bound to Fre at the beginning of the experiment. As shown in Figure 2A, addition of NADPH caused a decrease of absorbance corresponding to reduction of Rf. An isosbestic point at 336 nm was observed until the concentration of added NADPH became greater than that of Rf. In Figure 2C, a plot of the fractional absorbance changes at 446 nm as a function of the [NADPH]/[Rf] ratio showed that 1 mol of NADPH was sufficient to reduce 1 mol of Rf. Interestingly, close inspection of the absorbance spectra recorded throughout the experiment revealed that a weak long-wavelength absorption band developed with an isosbestic point at 528 nm (Figure 2B). Such an absorbance band could be assigned to a charge-transfer complex between reduced Rf and $NADP^+$ within Fre active site rather than to a flavin neutral semiquinone species. This conclusion could be reached on the basis of its shape (*33-37*), on the fact that it was observed at early stages of the titration when added NADPH had been totally converted to $NADP^+$, and that it persisted until the end of the titration.

In order to confirm this conclusion, the two following experiments were carried out. In one experiment, a solution containing Fre (30 μM), Rf (12.1 μM), $NADP^+$ (100 μM) and glucose 6-phosphate (2 mM) in buffer A, was made anaerobic. The absorbance spectrum of the solution was recorded before and immediately after the addition of glucose 6-phosphate dehydrogenase (1 U) from the side arm of the anaerobic cuvette. The spectrum of fully reduced flavin and of excess NADPH was obtained, which showed no absorption at wavelengths above 530 nm (data not shown).

In a second experiment, a solution containing Fre (70 μM), Rf (22.6

µM), NADP⁺ (166 µM), EDTA (5 mM) and deazaflavin (0.9 µM) in buffer A, was made anaerobic in the dark. Absorption spectra of the solution were recorded after irradiation for different intervals of time. As shown in Figure 3, absorbance changes consistent with reduction of Rf by photoreduced 5-deazaflavin were observed. A weak absorbance band developed at wavelengths greater than 528 nm (inset of Figure 3), which was similar to that observed during reaction of Fre with NADPH and Rf (Figure 2B). No long wavelength absorbance band developed in a similar photoreduction experiment in the absence of NADP⁺ (data not shown). It is noteworthy that in this experiment NADP⁺ was not reduced as no significant increase of the absorbance at 340 nm was observed. This result indicates that the reaction catalyzed by Fre is essentially irreversible.

Altogether, these data confirm that the observed long-wavelength absorbance band could be assigned to a charge-transfer complex between enzyme-bound reduced Rf and NADP⁺. It is likely that the weak intensity of the band reflects the fact that only a small amount of the charge-transfer complex is present as a consequence of its tendency to dissociate to yield the Fre.NADP⁺.Rf Michaelis complex and the free species, Fre, reduced Rf and NADP⁺. The latter hypothesis is consistent with the high $K_i$ value for NADP⁺ (5 mM, *14*).

*Kinetics of the Fre-catalyzed reaction between NADPH and Rf.*

The equilibrium spectrophotometric experiments presented above demonstrated that Fre stabilizes, although to a small extent, a charge-transfer complex between reduced Rf and NADP$^+$. This finding is in agreement with the hypothesis that Fre may be mechanistically similar to enzymes of the FNR family. The latter enzymes indeed form a charge-transfer intermediate between bound reduced flavin and oxidized pyridine nucleotide (CT-2 of Scheme 1). We wished to address the question of the ability of Fre to also stabilize a charge-transfer complex between bound Rf and NADPH, corresponding to CT-1 of Scheme 1, and to obtain indications of the possible participation of the charge-transfer complex intermediates to the catalytic turnover of Fre. Towards this goal, we used a rapid reaction (stopped-flow) spectrophotometer to monitor the reaction between Rf and NADPH catalyzed by Fre. It should be kept in mind that there are several limitations to the use of the rapid reaction approach to the study of the reaction of Fre. In particular, the amount of Fre that could be prepared and the limited solubility of Rf, together with its high extinction coefficient in the visible region, led us to adopt the experimental conditions outlined below. However, under our experimental conditions, given the properties of Fre ( e.g.: the relatively low affinity for the NADPH and Rf substrates), the time-course of absorbance changes observed at various wavelengths could not be used to calculate the rate constants associated with individual reaction steps. A much more complex set of measurements would have indeed be enquired in order to attempt a detailed kinetic analysis of Fre reaction, which was outside the scope of the present work.

Preliminary experiments were needed to set up the experimental design used throughout. In order to avoid the chemical reduction of Rf by NADPH, Fre was made anaerobic in the presence of a given concentration of Rf in a tonometer, while NADPH solutions were made anaerobic in glass syringes. In order to avoid photoreduction of Rf and significant temperature-dependent pH changes, the Tris/HCl buffer usually employed during kinetic studies (*14*) was substituted by Hepes/KOH buffer (buffer A). Finally, the steady-state kinetic parameters of the reaction were determined at various temperatures in Hepes/KOH and Tris/HCl buffers (Table 1) in order to detect any buffer effect on the properties of Fre, and to determine the turnover number of Fre ($k_{cat}$) and the $K_m$ value for NADPH under buffer and temperature conditions similar to those used during the stopped-flow experiments. No differences in the time-course of absorbance changes at various wavelengths were observed in the stopped-flow apparatus using different buffers or different combinations of the reagents to be mixed (namely, Fre versus a solution containing Rf and NADPH, as opposed to mixing a solution containing Fre and Rf with the NADPH solutions). Therefore, equilibration of Fre with NADPH and Rf to form the ternary Fre.NADPH.Rf Michaelis complex is most likely very rapid on the time-scale of the experiments.

A series of experiments was carried out at 25 °C. Equal concentrations of

Fre and Rf (32.7 µM after mixing) were reacted with NADPH (94, 200 and 365 µM, after mixing) and the time-course of absorbance changes was monitored at various wavelengths. As expected, absorbance at 450 nm decreased indicating Rf reduction (Figure 4A), but the time-dependence of such changes could not be described by functions corresponding to the sum of exponential processes. Rather, the initial (< 5ms) absorbance decrease at 450 nm was found to be linear with time (data not shown). This observation suggested that, rapidly, and before any significant amount of substrates were used, a state approximating steady-state turnover may be reached in solution. Indeed, by making the simple minded calculation of the initial reaction velocity of Rf reduction from the initial linear part of the traces observed at various NADPH concentrations, we calculated values that were consistent with an apparent $k_{cat}$ of 67 s$^{-1}$ and an apparent $K_m$ for NADPH of 21 µM. These values are very similar to those estimated from actual initial velocity measurements under steady-state conditions (Table 1).

At all NADPH concentrations tested, absorbance changes in the 520-750 nm range were qualitatively similar to each other. Figure 4A shows a representative trace of the time-course of absorbance changes observed at 550 nm. At all wavelengths between 520 and 750 nm, and at all NADPH concentrations used, only two distinct phases could be observed since a very rapid undetectable primary absorbance increase was completed within the instrument dead-time (1.7 msec) (The initial spectrum has no absorption at wavelengths between 520 and 750 nm). The first detectable phase is characterized by a slight absorbance increase. In the second phase, the absorbance decreases back to the baseline level. The absorbance at 550 nm reached a maximum while Rf was mainly present in the oxidized form, as indicated by the trace at 450 nm, and decayed as Rf got reduced (Figure 4A).

A good fit of the traces observed between 520 and 750 nm could be obtained using Eq. 3 and the same set of rates, $k_1$ and $k_2$.

$$A = A_1 \exp(-k_1 *t) + A_2 \exp(-k_2 *t) + C \qquad (3)$$

where : A, absorbance at time t; $A_1$ and $A_2$, pre-exponential terms which describe the amplitude of each phase; C, absorbance at the end of the reaction. At the highest NADPH concentration (365 µM), the first detectable phase of the reaction, which was associated with a small absorbance increase, took place with an apparent rate of 170 s$^{-1}$. The subsequent absorbance decrease appeared to occur with an apparent rate of 105 s$^{-1}$. Although it should be kept in mind that the calculated values cannot be strictly identified with rate constants that govern the individual steps of the reaction, it is interesting to note that the values we calculated are both significantly greater than the $k_{cat}$ value calculated for Fre reaction under steady-state conditions at 25 °C ( 75 □ 1 s$^{-1}$, Table 1). Therefore the hypothesis could be made that the observed transient species with long wavelength absorbance (namely, that formed within the instrument dead-time, 1.7 ms, and that accumulated at later stages of the reaction) could be part of the catalytic cycle.

In order to attempt a more detailed characterization of such intermediate species, the rapid kinetics experiments were also carried out at 5 °C. In a first series of experiments, equal volumes of an anaerobic equimolar mixture of Fre and Rf (40 μM each, after mixing) and of anaerobic NADPH solutions (90 μM, 165 μM, 197 μM or 378 μM, after mixing) were mixed in the stopped-flow apparatus.

The time-course of absorbance changes at various wavelengths were qualitatively similar to those observed during the experiments carried out at 25 °C (Figure 4B). The initial absorbance changes at 450 nm were linear with respect to time (data not shown). As noticed at 25 °C, by making the simple minded calculation of the initial reaction velocity of Rf reduction from the initial linear part of the traces observed at various NADPH concentrations, we calculated values that were consistent with an apparent $k_{cat}$ of 32 s$^{-1}$ and an apparent $K_m$ for NADPH of 22 μM. These values are very similar to those estimated from actual initial velocity measurements under steady-state conditions (Table 1).

At all NADPH concentrations, and at wavelengths between 520 and 800 nm, a rapid initial absorbance increase was now detectable, followed by a further slower increase and a final absorbance decrease back to the baseline level. All traces obtained at various wavelengths and various NADPH concentrations could be fitted with Eq. 4, which describes the sum of three exponential processes (Figure 4B).

$$A = A_1 \exp(-k_1 *t) + A_2 \exp(-k_2 *t) + A_3 \exp(-k_3 *t) + C \quad (4)$$

where : A, absorbance; $A_1$, $A_2$, $A_3$, pre-exponential terms which describe the amplitude of each phase; $k_1$, $k_2$, $k_3$, apparent rate constants; C, absorbance at the end of the reaction.

The values calculated for the observed second and third reaction phase ($k_2$, 83 □ 9 $s^{-1}$ and $k_3$, 38 □ 2 $s^{-1}$) were essentially independent from NADPH concentration (Figure 5). On the contrary, the calculated value of the observed rate associated with the initial rapid reaction phase increased with increasing NADPH concentration. Interestingly, the values of the observed $k_1$ appeared to show a hyperbolic dependence on NADPH concentration, and a value of 1247 □ 232 $s^{-1}$ could be extrapolated at infinite NADPH concentration from a double reciprocal plot (Figure 5). An apparent $K_m$ value for NADPH of 199 □ 72 µM was determined. However the physical meaning of this parameter cannot be determined at this stage.

The dependence of the observed rate describing the initial reaction phase on NADPH concentration is indeed that expected if this rate reflected that of formation of a charge-transfer complex between Fre-bound Rf and NADPH (analogous to CT-1 of Scheme 1) from the ternary Fre.NADPH.Rf Michaelis complex (corresponding to MC-1 of Scheme 1). Furthermore, the first reaction phase is significantly faster than the following reaction phases, which, in turn, both appear to be associated with rates greater than that of $k_{cat}$ determined for Fre during steady-state kinetic measurements at 5 °C. Finally, it should be noted that the absorbances of the reaction mixture at 550 nm and 700 nm were maximal while Rf was mainly present in the oxidized form, as indicated by the trace at 450 nm (Figure 4B).

Altogether, these results can be interpreted in terms of two transient reaction intermediates, whose absorbance at long wavelengths suggests that they may be charge-transfer complexes similar to CT-1 and CT-2 of Scheme 1. On the basis of the observed rates of appearance and decay of such transient species, the proposal that they may be part of the catalytic turnover of Fre can be made. In particular, if the hypothesis were made that the observed rates reflected the rates of conversion of MC-1 into CT-1 ($k_1$), CT-1 into CT-2 ($k_2$) and of CT-2 into MC-2 and free reaction products ($k_3$), it would appear that Fre catalytic turnover is determined by decay of CT-2.

*Absorbance spectra of reaction intermediates.*

As expected from the results presented above, the long-wavelength absorbing species were found to accumulate to a greater extent during rapid kinetic experiments performed in the presence of higher concentrations of Rf with respect to Fre (37 μM Fre, 147 μM Rf and 400 μM NADPH). From inspection of traces obtained at several wavelengths between 520 and 800 nm, it was clear that two intermediates were formed. The relative amounts of these species reached a maximum at 3 and 75 ms of reaction, respectively. As shown in Figure 6, at early stages of the reaction (up to 3 ms), only 2 % of Rf was reduced and a species with high absorbance at 550 nm and low absorbance at 750 nm was formed. As more Rf was reduced, another species characterized by a lower absorbance at 550 nm, but a relatively higher absorbance at 750 nm accumulated (maximum at 75 ms). Absorbance changes observed at various wavelengths between 520 and 800 nm could be fitted with Eq. 5, which describes the sum of four exponential processes using the same set of rates , $k_1 = 1046$ s$^{-1}$, $k_2 = 52$ s$^{-1}$, $k_3 = 31$ s$^{-1}$, $k_4 = 9.6$ s$^{-1}$ (inset of Figure 6).

$$A = A_1 \exp(-k_1 *t) + A_2 \exp(-k_2 *t) + A_3 \exp(-k_3 *t) + A_4 \exp(-k_4 *t) + C \qquad (5)$$

where : A, absorbance; $A_1$, $A_2$, $A_3$, $A_4$, pre-exponential terms, which describe the amplitude of each phase; $k_1$, $k_2$, $k_3$, $k_4$, apparent rate constants; C, absorbance at the end of the reaction. Thus, from the function describing the time-course of the reaction at the various wavelengths, it was possible to calculate the absorbance values of the reaction mixture at different wavelengths and different reaction times. Figure 7 shows the spectra of the reaction mixture calculated at 3 and 75 ms, when the amounts of the first and the second intermediate, respectively, were maximal, and at 1 s, when the reaction was completed. The spectrum of the starting solution containing Fre (37 µM) and Rf (147 µM), which was recorded with the rapid scanning device of the stopped-flow apparatus, is also shown in the figure.

The calculated spectra can be compared with those obtained during studies of a variety of flavoproteins (*33-37*), especially those from the FNR family (*23, 25-28*). Recently, the reaction with NADPH of the truncated form of PDR, lacking the ferredoxin-like domain (PDR-FeS), was studied in detail (*26*) and spectra of the charge-transfer complex intermediates were deconvoluted from rapid reaction data. The absorbance spectra calculated during the reaction of NADPH and Rf catalyzed by Fre (Figure 7) at 3 and 75 ms present features similar to those of CT-1 and CT-2 of PDR-FeS, respectively (see Figure 10C of reference *26*). This observation strengthens the hypothesis that the first intermediate, formed when Rf is still mostly in the oxidized form, is a charge-transfer complex between Fre-bound oxidized Rf and NADPH, while the second intermediate, which accumulates at later stages of the reaction, is a charge-transfer complex between Fre-bound reduced Rf and $NADP^+$.

However, in our case, such intermediate is unstable and decays. Interestingly, a low amount of $(Fre.Rf_{red}.NADP^+)^*$ charge-transfer complex was still present at the end of the reaction (Figure 7, spectrum at 1 s) in agreement with the results of the static spectrophotometric experiments

(Figures 2B and 3). In the case of PDR-FeS (*26*) an extinction coefficient of 3200 M$^{-1}$ cm$^{-1}$ at 725 nm was determined for the charge-transfer complex CT-2 formed between reduced flavin and oxidized pyridine nucleotide. Using this value, we can roughly estimate that about 6% of Fre is present as (Fre.Rf$_{red}$.NADP$^+$)* charge-transfer complex at the end of the reaction (Figure 7). NADP$^+$, which is an inhibitor of Fre reaction, competitive with respect to NADPH, exhibits a very low affinity for the enzyme, with a $K_i$ value of 5 mM (*14*). At the end of the experiment (Figures 6 and 7) concentrations of the different species were the following: 37 µM Fre, 147 µM reduced Rf, 147 µM NADP$^+$ and 253 µM NADPH. Under these conditions, if we consider that the presence of reduced Rf and NADPH does not affect the $K_i$ value for NADP$^+$, we can calculate that 3% of NADP$^+$ may be bound to Fre. This value is in line with that calculated above for the (Fre.Rf$_{red}$.NADP$^+$)* charge-transfer complex present at the end of the experiment shown in Figure 7.

CONCLUSIONS

The reduction of Rf by NADPH catalyzed by Fre was studied by static and rapid kinetics absorption spectroscopies at 25 and 5 °C. The data reported here demonstrate the occurrence of two distinct intermediates during the catalytic cycle and allow us to propose a minimal mechanism for the reaction (Scheme 2). These intermediates, which are better detected at low temperature, exhibit spectroscopic properties similar to those identified as CT-1 and CT-2 during reaction of enzymes of the FNR family (Scheme 1, *23, 25-28*). In particular, they show significant absorption at long wavelengths, which is characteristic of charge-transfer transitions between the nicotinamide ring and the flavin isoalloxazine.

As outlined in Scheme 2, it can be proposed that the initial reaction phase we observed in the stopped-flow most likely includes 3 steps : (a) binding of NADPH and of (b) Rf to form the Fre.Rf.NADPH Michaelis complex, MC-1, followed by (c) conversion of MC-1 into the first charge-transfer complex, CT-1, between oxidized Rf and NADPH, (Fre.Rf.NADPH)*. Free Rf, Fre.Rf and MC-1 all have no absorbance at wavelengths greater than 520 nm and cannot account for the long-wavelength absorption bands detected during the reaction. During the second phase detected in our experiments, hydride transfer may take place to yield a (Fre.Rf$_{red}$.NADP$^+$)* charge-transfer complex, CT-2, between reduced Rf and

NADP$^+$. The latter would decay in the last observed reaction phase to give the Fre.Rf$_{red}$.NADP$^+$ Michaelis complex (MC-2) which would dissociate to yield free Fre, reduced Rf and NADP$^+$. Again, the spectral properties of free Rf, enzyme-bound reduced Rf and MC-2 would not be detectable at long-wavelength and cannot account for the absorption bands detected during the reaction.

The results presented here support our initial hypothesis of a similarity of Fre with enzymes of the FNR family. Such hypothesis was first formulated on the basis of limited sequence similarities and supported by site-directed mutagenesis studies of the enzyme (*24*). The present study extends the similarity of Fre to members of the FNR family at the functional level. Thus, although Fre uses the flavin as a substrate rather than as a prosthetic group, the reduction of Rf by NADPH within the enzyme active center may occur through the same mechanism as that used by enzymes of the FNR family (*23, 25-28*).

ACKNOWLEDGMENTS

Table 1. Comparison of the values of $k_{cat}$ and $K_m$ for NADPH obtained from steady-state kinetic measurements at 25 and 5 °C.

| Temperature (°C) | $k_{cat}$ (s$^{-1}$) | $K_m$ (NADPH) (µM) |
|:---:|:---:|:---:|
| 25 | 78.3±3[a] | 32 ± 2[a] |
| 25 | 75 ± 1[b] | 29 ± 1[b] |
| 5 | 27 ± 2[a] | 31 ± 2[a] |
| 5 | 26 ± 2[b] | 19 ± 2[b] |

Steady-state kinetic measurements were carried out as described in Materials and Methods in [a] 50 mM Tris/HCl buffer, pH 7.6 or [b] 50 mM Hepes/KOH buffer, pH 7.6, 10% glycerol (buffer A).

FIGURE LEGENDS

Figure 1 : Spectral properties of the Fre-Rf complex. Aliquots of a 521 µM solution of Fre were added to a solution containing 11.7 µM Rf in buffer A, at 10 °C. Absorbance changes were complete immediately after each Fre addition. The spectra, from top to bottom at 446 nm, are those recorded after addition of 0, 0.37, 0.75, 1.12, 1.5, 1.87, 2.24 and 2.63-fold molar excess Fre. The inset shows the fractional changes of absorbance at 446 nm, after correction for dilution, plotted as a function of the [Fre]/[Rf] ratio calculated from the spectra shown in the main panel (open circles) and from a similar experiment in the presence of 166 µM AADP (closed circles). The curve corresponds to that calculated using Eq. 1 assuming $K_d$ = 3.6 µM and n = 1.16.

Figure 2 : NADPH reduction of Rf catalyzed by Fre. A solution containing 50 µM Rf and 175 µM Fre in buffer A was made anaerobic and maintained at 10 °C. Aliquots of a solution of NADPH (4.13 mM) were anaerobically added through a syringe fitted on the side arm of the anaerobic cuvette. Panel A shows, from top to the bottom, the spectra recorded after addition of 0, 0.17, 0.31, 0.46, 0.6, 0.74, 0.89, 1.03, 1.18 -fold molar excess of NADPH with respect to Rf. Panel B shows a blow-up of the spectra at long wavelengths. For sake of clarity, only spectra obtained after addition of 0, 0.17, 0.31, 1.18, 1.3-fold excess of NADPH are shown. Panel C shows the plot of fractional absorbance changes observed at 446 nm, as a function of the [NADPH]/[Rf] ratio.

Figure 3 : Photoreduction of Rf in the presence of $NADP^+$ and Fre. A solution containing 22.6 µM Rf, 70 µM Fre, 166 µM $NADP^+$, 5 mM EDTA, 0.9 µM 5-deazaflavin in buffer A was made anaerobic and maintained at 10 °C. The figure shows, from top to bottom, the spectra recorded after 0, 15, 30, 45, 65, 85, 105, 125 seconds of irradiation. The inset shows a blow-up of the spectra at long wavelengths. For sake of clarity, only spectra obtained after 0, 30, 60, 105 seconds of irradiation are presented.

Figure 4 : Time-course of absorbance changes observed during the reaction of Fre with Rf and NADPH at 25 and 5 °C. Panel A. Fre (32.7 µM) was reacted with Rf (32.7 µM) and NADPH (365 µM) in buffer A at 25°C. Absorbance changes were monitored at 450 (♦) and 550 (●) nm. The dashed line is the best fit of the trace observed at 550 nm to Eq. 3 assuming $k_1 = 170$ $s^{-1}$, $k_2 = 105$ $s^{-1}$, $A_1 = -0.037$, $A_2 = 0.056$, $C = 0.04$. In order to allow direct comparison of the data at 450 and 550 nm, the absorbance values at 550 nm and of the fitted curve were multiplied by a factor of 15, and a constant value was subtracted to superimpose the traces at long reaction time. Panel B. Fre (40 µM) was reacted with Rf (40 µM) and NADPH (378 µM) in buffer A at 5°C. The absorbance changes at 450 (♦), 550 (●) and 700 (■) nm are shown. The dashed lines are the best fit of the traces observed at 550 and 700 nm to Eq. 4 assuming in all cases $k_1 = 886$ s$^{-1}$, $k_2 = 81$ s$^{-1}$, $k_3 = 39$ s$^{-1}$. The best fit of the 550 nm trace was obtained with $A_1 = -0.020$, $A_2 = -0.033$, $A_3 = 0.071$, $C = 0.013$; for the 700 nm trace : $A_1 = -0.005$, $A_2 = -0.012$, $A_3 = 0.013$, $C = 0.016$. In order to allow direct comparison of the traces obtained at various wavelengths, the absorbance changes observed at 550 and 700 nm, along with the corresponding calculated curves were multiplied by a factor of 10 and 30 respectively. Constant values were also subtracted to superimpose them to the 450 nm trace at long reaction times.

Figure 5 : Dependence on NADPH concentration of the observed rates of reaction of Fre with Rf and NADPH. Fre (40 µM) was reacted with Rf (40 µM) and various NADPH concentrations (378, 227, 197, 165, 90 µM) in buffer A at 5 °C in a stopped-flow apparatus under anaerobiosis. Traces obtained at 550, 580, 600 and 650 nm with different NADPH concentrations are fitted according to Eq. 4 . The figure shows the plots of the reciprocal values of the observed rate constants, $k_1$ (o), $k_2$ (Δ) and $k_3$ (□), defined from Eq. 4, as a function of the reciprocal values of the corresponding NADPH concentration. The lines shown are the best fits of the data between 378 and

165 µM NADPH concentrations, assuming: maximum value of observed $k_1 = 1247$ s$^{-1}$ and apparent $K_{m\ (NADPH)} = 199$ µM; $k_2 = 83$ s$^{-1}$ and $k_3 = 38$ s$^{-1}$.

Figure 6 : Time-course of the absorbance changes observed during the reaction of Fre with excess Rf and NADPH at 5 °C. Fre (37 µM) was reacted with Rf (147 µM) and NADPH (400 µM) in buffer A at 5°C. Absorbance changes were monitored at 450 (♦), 550 (●) and 750 (■) nm. In order to allow direct comparison of the traces obtained at various wavelengths, the absorbance change observed at 550 and 750 nm, along with the corresponding calculated curves (inset) were multiplied by a factor of 20 and 40 respectively. Constant values were also subtracted to superimpose them to the 450 nm trace at long reaction times. In the inset, the data are shown using a logarithmic time-scale. The dashed lines are the best fits of the traces observed at 550 and 750 nm to Eq. 5 assuming in all case $k_1 = 1046$ s$^{-1}$, $k_2 = 52$ s$^{-1}$, $k_3 = 31$ s$^{-1}$, $k_4 = 9.6$ s$^{-1}$. Best fit of the 550 nm trace was obtained with $A_1 = -0.022$, $A_2 = 0.066$, $A_3 = -0.10$, $A_4 = 0.075$, $C = 0.026$; for the 750 nm trace: $A_1 = -0.004$, $A_2 = 0.025$, $A_3 = -0.048$, $A_4 = 0.026$, $C = 0.007$.

Figure 7 : Calculated spectra of the solution during the reaction of Fre with Rf and NADPH at different times after mixing in the stopped-flow spectrophotometer at 5 °C. Spectra after 3 ms (♦), 75 ms (■) and 1000 ms

(●) were calculated from the best fits of the traces observed at 525, 550, 575, 600, 625, 650, 675, 700, 725, 750, 775 and 800 nm to Eq. 5, using $k_1$ = 1046 s$^{-1}$, $k_2$ = 52 s$^{-1}$, $k_3$ = 31 s$^{-1}$, $k_4$ = 9.6 s$^{-1}$. The dashed line is the spectrum of Fre (37 µM) and Rf (147 µM) in buffer A.